\documentclass{PoS}

\title{Indirect dark matter search with the balloon-borne PEBS detector}

\ShortTitle{PEBS}

\author{\speaker{Henning~Gast}, Roman~Greim, Thomas~Kirn, Gregorio~Roper Yearwood, and Stefan~Schael\\
        I.~Physikalisches Institut B, RWTH Aachen, Aachen, Germany\\
        E-mail: \email{henning@physik.rwth-aachen.de}\\
		\email{greim@physik.rwth-aachen.de}\\
		\email{kirn@physik.rwth-aachen.de}\\
		\email{roper@physik.rwth-aachen.de}\\
		\email{schael@physik.rwth-aachen.de}}

\abstract{A precision measurement of the cosmic-ray positron spectrum may help
to solve the puzzle of the nature of dark matter. Pairwise
annihilation of neutralinos, predicted by some supersymmetric
extensions to the standard model of particle physics, may leave a
distinct feature in the cosmic-ray positron spectrum.

As the available data are limited both in terms of statistics and
energy range, we are developing a balloon-borne detector (PEBS) with a
large acceptance of $4000\,\mathrm{cm}^2\,\mathrm{sr}$.
A superconducting magnet creating a field of $0.8\,\mathrm{T}$ and a tracking
device consisting of scintillating fibers of $250\,\mu\mathrm{m}$ diameter with
silicon photomultiplier readout will allow rigidity and charge
determination to energies above $100\,\mathrm{GeV}$. The dominant proton background is
suppressed by the combination of an electromagnetic calorimeter and a
transition radiation detector consisting of fleece
layers interspersed with straw-tube proportional counters. The
calorimeter uses a sandwich of tungsten and scintillating fibers that
are again read out by silicon photomultipliers.

The design study, based on a detailed Geant4 simulation and testbeam
measurements, will be presented along with an interpretation of the
currently available positron data in the context of the mSUGRA
model. The constraints that future precise measurements could put on
this model will be discussed.}

\FullConference{Identification of dark matter 2008\\
		 August 18-22, 2008\\
		 Stockholm, Sweden}

\begin{document}
\section{Detector design}

As an experiment designed to measure the positron component in the cosmic
rays with high precision, the Positron-Electron-Balloon-Spectrometer
(PEBS) has to meet several crucial requirements: First of all, the geometric
acceptance needs to be large due
to the small flux of positrons. A clean positron sample can only be
obtained if a suppression of the predominant proton background
on the level of one in one million is achieved. In addition,
a good momentum resolution is necessary for charge sign
determination and subsequent electron suppression.
The design study presented here is based on a full simulation of the
behavior of the experiment using
the Geant4 package\cite{ref:g4}. In addition, key elements have
been verified in a series of testbeams at CERN over the years 2006-2008.\\
\par
\begin{figure}[htb]
\begin{center}
\begin{tabular}{cc}
\includegraphics[width=0.5\textwidth,angle=0]{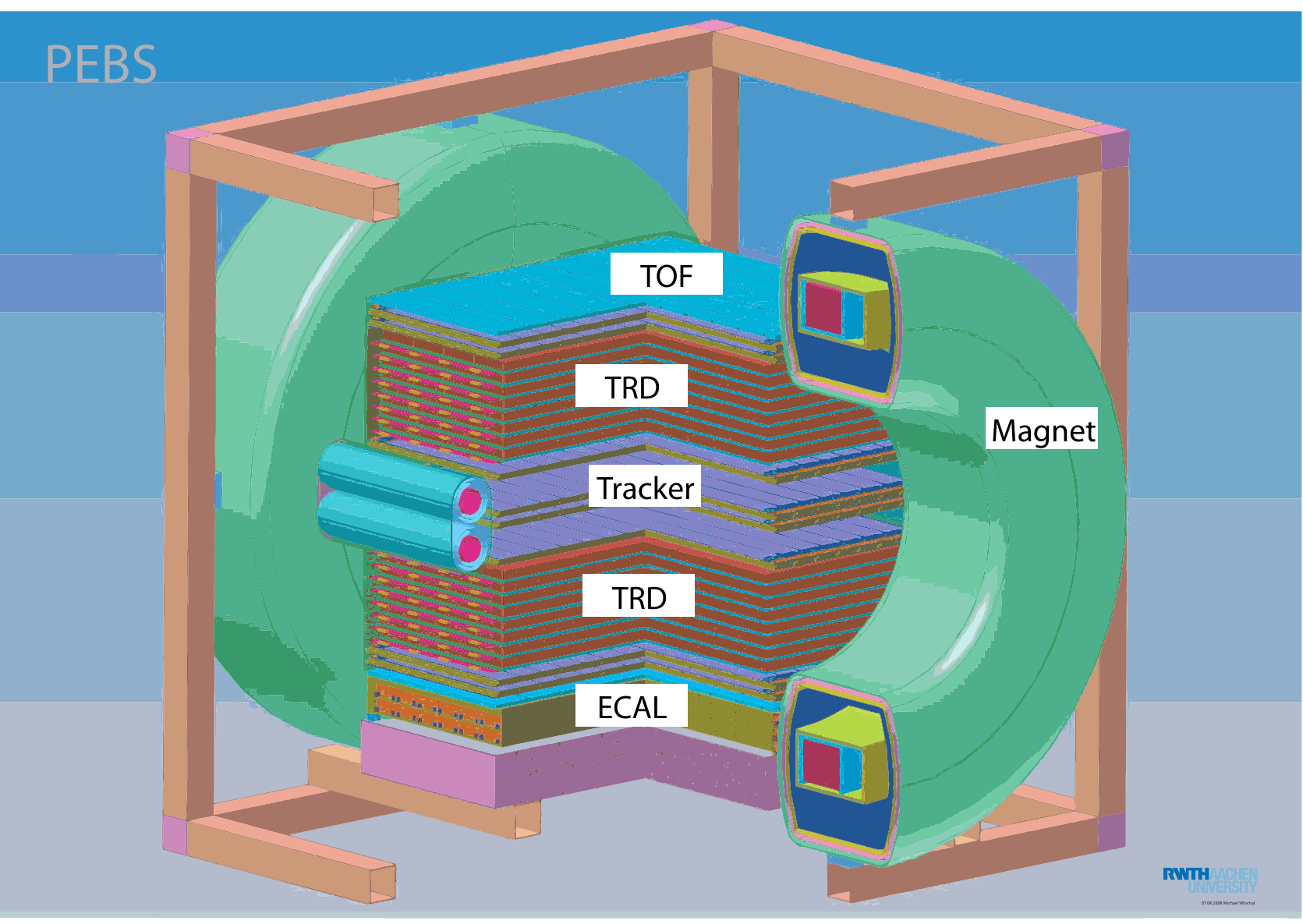}&
\includegraphics[width=0.4\textwidth,angle=0]{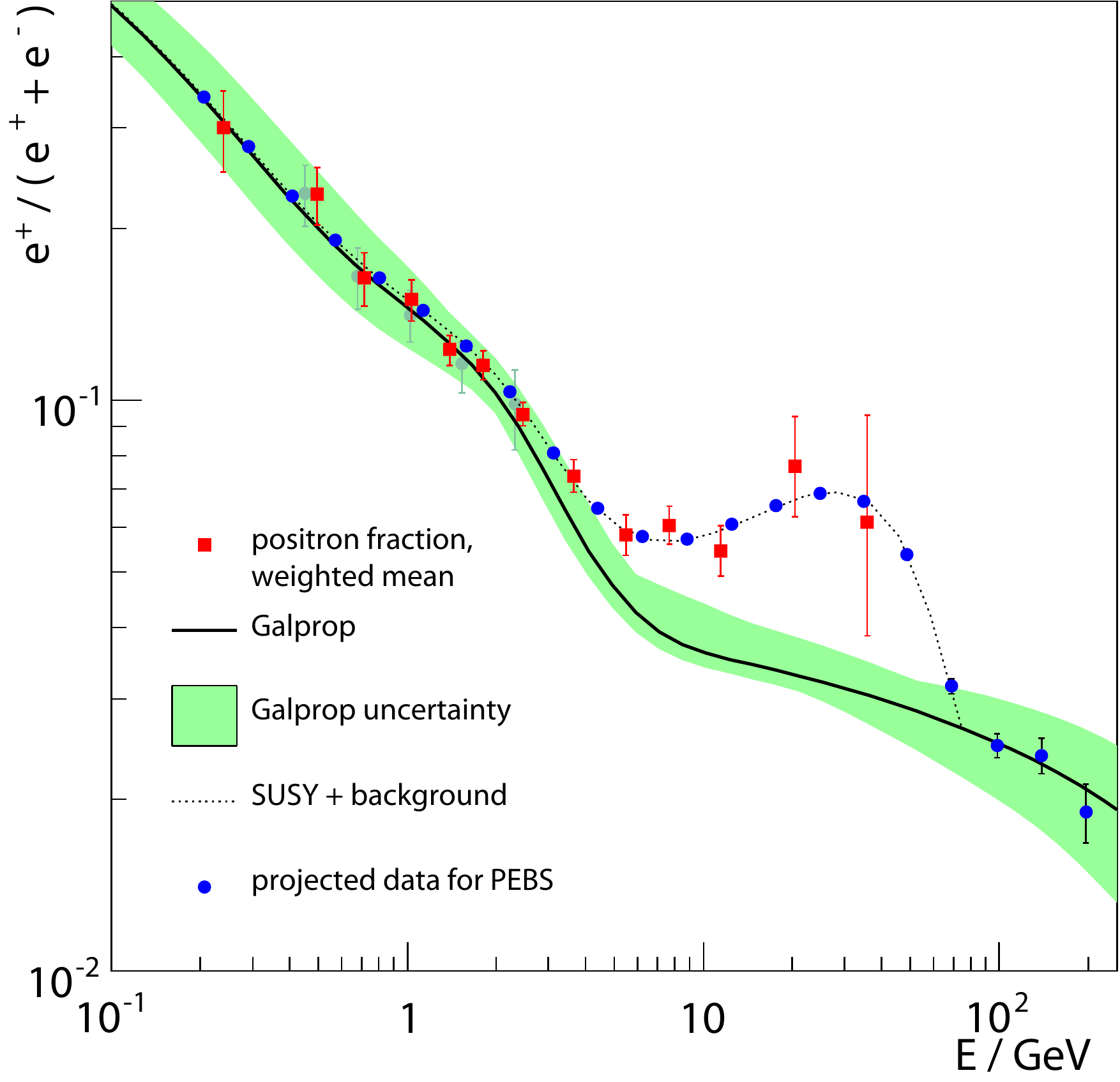}\\
\end{tabular}
\end{center}
\caption{{\it Left:} Cut-out mechanical drawing of the PEBS detector.
{\it Right:} Weighted mean of the positron fraction data from 
  AMS-01\cite{ref:ams01old,ref:ams01new}, HEAT\cite{ref:heat},
  CAPRICE\cite{ref:caprice}, and TS93\cite{ref:ts93}, together with the secondary background as
  predicted by the Galprop\cite{ref:galproprev} conventional model. The uncertainty band for the Galprop model was
obtained from a variation of model parameters within the bounds
allowed by the B/C data and the positron fraction data below
$3\,\mathrm{GeV}$. The statistical uncertainties achievable with a
detector that has a geometric acceptance of
$0.4\,\mathrm{m}^2\,\mathrm{sr}$ and measures for 100~days are
projected for the positron and electron fluxes of a dark matter
scenario with an mSUGRA neutralino ($m_0=1560\,\mathrm{GeV}$,
$m_{1/2}=260\,\mathrm{GeV}$, $\tan\beta=40$, $A_0=0$ and $\mathrm{sgn}\mu=+1$) on top of
the secondary backgrounds. A boost factor of~150 is used for the
signal fluxes, chosen to give the best fit to the presently available
data. DarkSUSY4.1\cite{ref:darksusy} has been used for the calculation
of the signal fluxes.}
\label{fig1}
\end{figure}

PEBS (fig.~\ref{fig1} ({\it left})) has been designed to meet the
requirements stated above.
A magnetic field of mean flux density $B=0.8\,\mathrm{T}$ and mean $BL^2=0.62\,\mathrm{Tm}^2$ is created by two
superconducting Helmholtz coils, located inside a helium cryostat.
The curvature of a charged particle's trajectory in this field is measured by
a scintillating fiber tracker with silicon photomultiplier readout\cite{ref:scifitracker}. A
transition radiation detector (TRD), located between the tracker
super-layers, and an electromagnetic calorimeter (ECAL) at the bottom of the
experiment provide rejection power against protons. 
Scintillator panels above and below the tracker act as a
time-of-flight system (TOF) and are used for triggering purposes.

Earth's atmosphere prohibits a measurement of GeV-range cosmic rays
on the ground. While space experiments have the undisputed virtue of being able to
measure the spectra of cosmic rays completely undisturbed by the
atmosphere, scientific high-altitude balloons constitute an
interesting alternative for several reasons. The experiment can be salvaged after the flight
and be recalibrated, refitted and eventually repeated for the gradual
improvement of the statistical accuracy, and it can be
conducted at a much lower cost. Mission durations of up to
40~days have been reached by
traveling with the circular arctic winds around the North or South
Pole\cite{ref:cream}. The amount of helium carried for cooling the superconducting
magnet will be sufficient for such a measurement time.

The geometric acceptance of the detector is limited by the weight
and power constraints imposed by the carrier system. The most
important contributions to the overall weight are the
magnet weight and the weight of the calorimeter with $850\,\mathrm{kg}$ and
$600\,\mathrm{kg}$, respectively. The power consumption is dominated by the
$300\,\mathrm{W}$ needed for the tracker which has roughly $55\,000$ individual
readout channels.

The tracker has an aperture of $86\times86\,\mathrm{cm}^2$ and a
length of $80.6\,\mathrm{cm}$. From
the positions and dimensions of all the sensitive areas of the
subdetectors, the geometric acceptance of PEBS is calculated to be
$0.4\,\mathrm{m}^2\,\mathrm{sr}$.\\
\par
The tracking device will consist of scintillating fibers grouped into
modules and read out by linear silicon photomultiplier arrays (SiPMs). A module
comprises two stacks of round fibers of $250\,\mu\mathrm{m}$ diameter,
128~fibers wide and five fibers high, glued together in the tightest arrangement. The
stacks are held apart by two carbon fiber skins with Rohacell foam in
between. Using
scintillating fibers, the material budget in the particles' flight
path through the tracker does not exceed
$6\,\%$ of a radiation length, while the TRD will contribute another
$6\,\%$. The modules will be grouped into eight layers, two of those
being located at the entrance and exit of the tracking device,
respectively, and four in the center.

Silicon photomultipliers\cite{ref:sipm1} have the virtues of being insensitive to
magnetic fields, having high quantum efficiency, as well as
compactness and auto-calibration. They will therefore be used to
detect the photons trapped in the scintillating fibers and will be
read out by a dedicated VA chip. Linear arrays containing 32~silicon
photomultiplier columns each are located at alternating ends of the fiber
bundles. The remaining end of each fiber is covered by a reflective coating to
increase the light yield by a factor of roughly $1.6$. Five
fibers in one column are then optically connected to one SiPM column.
The weighted cluster mean from amplitudes in adjacent SiPMs columns will be
calculated to pinpoint the intersection of a trajectory with a fiber
module.

A prototype of a tracker module has been subjected to a $10\,\mathrm{GeV}$ proton testbeam at the
CERN~T9 beamline. The measured light yield of 11~photo electrons per
MIP crossing was used as input to the PEBS Monte Carlo simulation. A
momentum resolution of $16\,\%$ is predicted for $100\,\mathrm{GeV}$ protons
in this case.\\
\par
A sandwich calorimeter for three-dimensional shower reconstruction has been designed
to provide rejection power against the predominant proton component in
the cosmic rays. It comprises 20~layers consisting of $2\,\mathrm{mm}$ tungsten 
interleaved with layers of scintillator
bars of $2\,\mathrm{mm}$ height and $7.75\,\mathrm{mm}$ width. They are read out 
by individual SiPMs of $1\times1\,\mathrm{mm}^2$ area which
are sitting in front of wavelength-shifting fibers embedded in the
scintillator bars on both ends. Attenuation is used on one side to increase the dynamic range. Five layers are grouped
into a super-layer and four super-layers are placed with alternating
direction.
The total depth of the calorimeter is $11.4$~radiation lengths.

A preliminary cut-based analysis, using the PEBS Geant4 simulation, has been performed to study the proton rejection of
this setup. For each event, a shower fit using a standard Gamma
function parameterization has been performed and the following
variables have been used to distinguish 
positrons from protons: $E/p$-match, total shower amplitude,
fitted shower maximum, ratio of shower energy within one Moli\`ere radius from the
shower axis and angle between the reconstructed track and shower axis.

Proton rejections of the order of 1000 can
easily be achieved already with this rather
coarse method. The corresponding electron efficiency is around $75\,\%$.\\
\par
The design of the transition radiation detector is based on the one
constructed for the AMS-02 experiment on the International Space
Station\cite{ref:ams02trd}. The TR x-ray photons are generated in a
$2\,\mathrm{cm}$ thick irregular fleece radiator made of polyethylene and
polypropylene. They are subsequently detected in proportional wire
chambers in the form of straw tubes made of
aluminized kapton foils which have an inner diameter of $6\,\mathrm{mm}$ and
are filled with an $80:20$ mixture of Xe/CO$_2$. The straw tubes are
grouped into modules and eight layers each are placed in the gaps
above and below the central tracking layers. The proton rejection yielded by the TRD reaches a value of 1000 at
$80\,\%$ electron efficiency in the interesting energy range.

\section{Performance study}
Detailed performance studies using both Monte Carlo and testbeam data
have been conducted.
As the ECAL and the TRD measure independently, their combined proton
rejection power can be expected to be as high as one in one million.
The large acceptance, good momentum resolution and reliable proton
suppression of PEBS would allow a precision measurement of the
cosmic-ray positron fraction (fig.~\ref{fig1} ({\it right})) up to
energies above $100\,\mathrm{GeV}$.\\
\par
\begin{figure}[htb]
\begin{center}
\begin{tabular}{cc}
\includegraphics[width=0.5\textwidth,angle=0]{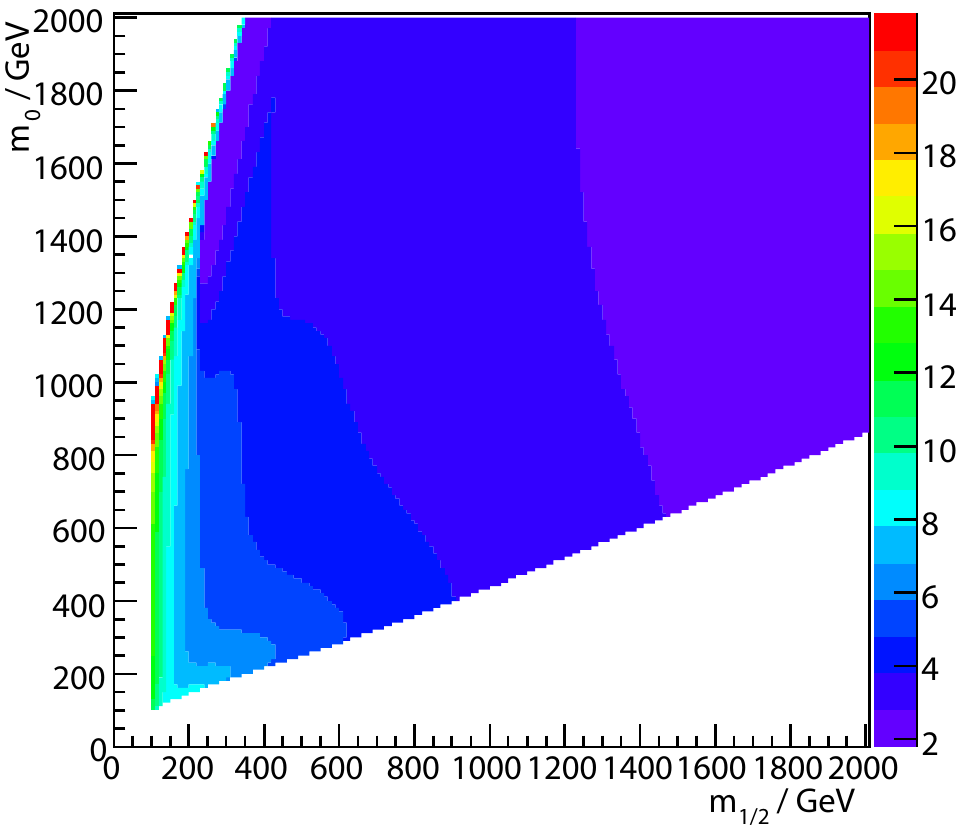}&
\includegraphics[width=0.46\textwidth,angle=0]{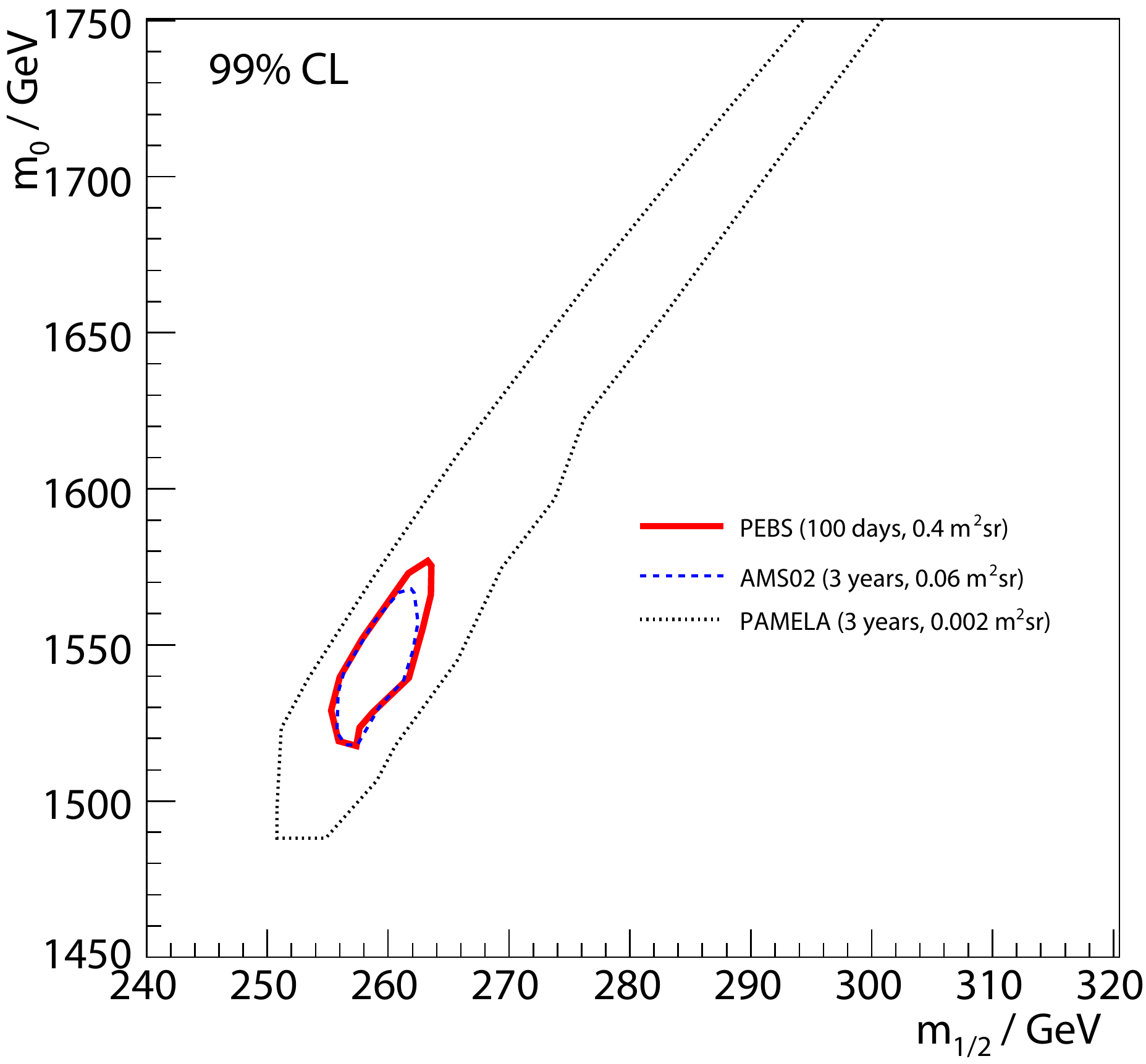}
\end{tabular}
\end{center}
\caption{{\it Left:} $\chi^2$-contour as described in the text, for the weighted
mean of the currently available positron fraction data. {\it Right:} Projected $99\,\%$ CL-contour 
for the statistical accuracy to be expected from PEBS, as compared to PAMELA and AMS-02,
for the benchmark point used in fig.~1 ({\it right}), in the
$m_{1/2}$-$m_0$-plane, for fixed $\tan\beta=40$. Note the different scales.}
\label{fig2}
\end{figure}

As an example for the physics performance to be expected from PEBS, a
scan of the mSUGRA parameter space was conducted. For the case that
mSUGRA is realized in nature and the neutralino contained in this
model constitutes the dark matter, the signal fluxes
$\Phi^\mathrm{sig}(e^\pm)$ for positrons and
electrons resulting from
neutralino annihilations in the Galactic halo were calculated using
DarkSUSY~4.1\cite{ref:darksusy}. 
The model fluxes were calculated as the sum of the background fluxes
$\Phi^\mathrm{bg}(e^\pm)$, taken from the conventional Galprop
model\cite{ref:galpropconventional}, and
the boosted signal fluxes:
\begin{equation}
\label{eq:boostedflux}
\Phi^\mathrm{model}(e^\pm)=\Phi^\mathrm{bg}(e^\pm)+f_b\cdot\Phi^\mathrm{sig}(e^\pm)
\end{equation}
The boost factor $f_b$ in equation (\ref{eq:boostedflux}) was used as
the only free parameter in a fit of the resulting
positron fraction for a given point in mSUGRA parameter space
to the weighted mean of the currently available data
(fig.~\ref{fig1}). The resulting contour of the best-fit $\chi^2$ is
essentially flat over vast amounts of the parameter space
(fig.~\ref{fig2} {\it left}). This situation can be expected to change
drastically with the arrival of data from PEBS. For the benchmark
model given in the caption of fig.~\ref{fig1}, random data for an
experiment with the acceptance of PEBS and a measurement time of
100~days were generated and the procedure described above was
repeated. Now, the $\chi^2$-contour (fig.~\ref{fig2} {\it right}) has
a distinct minimum at the benchmark point. In this model and at this level of
statistical accuracy, the corresponding
resolution of the neutralino mass would be limited by
the energy resolution of the calorimeter which is expected to be on
the order of $7\,\%$ at $100\,\mathrm{GeV}$.


\begin{thebibliography}{99}
\bibitem{ref:g4}
S.~Agostinelli et al., Nucl.~Instr.~Meth.~A {\bf 506} (2003) 250-303
\bibitem{ref:cream}
H.S.~Ahn et al., Nucl.~Instr.~Meth.~A {\bf 579} (2007) 1034-1053
\bibitem{ref:sipm1}
B.~Dolgoshein et al., Nucl.~Instr.~Meth. A {\bf 563} (2006), 368-376
\bibitem{ref:ams02trd}
P.~v.~Doetinchem et al., Nucl.~Instr.~Meth. A {\bf 558} (2006) 526-535
\bibitem{ref:ams01old}
J.~Alcaraz et al., Phys.~Lett.~B {\bf 484} (2000) 10-22
\bibitem{ref:ams01new}
M.~Aguilar et al., Phys.~Lett.~B {\bf 646} (2007) 145-154
\bibitem{ref:heat}
J.J.~Beatty et al., Phys.~Rev.~Lett. {\bf 93} (2004) 241102
\bibitem{ref:caprice}
M.~Boezio et al., ApJ {\bf 532} (2000) 653-669
\bibitem{ref:ts93}
R.L.~Golden et al., ApJ {\bf 457} (1996) L103-L106
\bibitem{ref:galproprev}
A.W.~Strong et al., Annu.~Rev.~Nucl.~Part.~Sci. {\bf 57} (2007) 285-327
\bibitem{ref:darksusy}
P.~Gondolo et al., JCAP 0407 (2004) 008
\bibitem{ref:scifitracker}
H.~Gast et al., Nucl.~Instr.~Meth. A {\bf 581} (2007) 423-426
\bibitem{ref:galpropconventional}
V.S.~Ptuskin et al., ApJ {\bf 642} (2006) 902-916

\end{thebibliography}
\end{document}